\documentclass[prb,
twocolumn,
superscriptaddress,showpacs,amsmath,amssymb]{revtex4}

\begin{document}

\title{Effect of the inner horizon on the black hole thermodynamics:
 Reissner-Nordstr\"om black hole and Kerr black hole}

\author{G.E.~Volovik}
\affiliation{Low Temperature Laboratory, Aalto University,  P.O. Box 15100, FI-00076 Aalto, Finland}
\affiliation{Landau Institute for Theoretical Physics, acad. Semyonov av., 1a, 142432,
Chernogolovka, Russia}

\date{\today}

\begin{abstract}
For the Schwarzschild black hole the Bekenstein-Hawking entropy is proportional to the area of the event horizon.
For the black holes with two horizons the thermodynamics is not very clear, since the role of the inner horizons is not well established. Here we calculate the entropy of the Reissner-Nordstr\"om black hole and of the Kerr black hole, which have two horizons.
For the spherically symmetric Reissner-Nordstr\"om black hole we used several different approaches. All of them give the same result for the entropy and for the corresponding temperature of the thermal Hawking radiation. The entropy is not determined by the area of the outer horizon, and it is not equal to the sum of the entropies of two horizons.  It is determined by the correlations between the two horizons, due to which the total entropy of the black hole and the temperature of Hawking radiation depend only on mass $M$ of the black hole and do not depend on the black hole charge $Q$. For the Kerr and Kerr-Newman black holes it is shown that their entropy has the similar property: it depends only on mass $M$ of the black hole and does not depend on the angular momentum $J$ and charge $Q$.
\end{abstract}
\pacs{
}

\maketitle

\section{Introduction}

For the simplest black hole the Bekenstein-Hawking entropy is proportional to the area of the event horizon, $S=A/4G$.
However, for the black hole with two or more horizons, such as the rotating and charged black holes and the black holes in the de Sitter space-time, the thermodynamics is not well established, since the role of the inner horizon remains unclear. 
The inner  horizon may experience instabilities,\cite{Penrose1968,McNamara1978,Matzner1979,Starobinsky1979,Poisson1990,Hod1998,Burko1998,Chesler2020}
which may or may not produce singularity at the horizon, or even prevent the formation of the inner horizon.

If the inner horizon is present it may contribute to the entropy of the black hole. However,
the total entropy of the system is not necessarily the sum of the entropies of the horizons, since the correlations between the horizons are possible, see e.g. Refs.\cite{YunHe2018,YuboMa2021}.  
There are the other hints, which suggest that the black hole entropy is not necessarily determined  by the area of the outer horizon, see e.g. extremal surfaces in Refs.\cite{Penington2020,Choudhury2021,Manu2021} and references therein.

Here we consider the charged black hole (the  Reissner-Nordstr\"om black hole) and the rotating black hole (the Kerr black hole). We ignore the possible instabilities of the inner horizons of these black holes. 
The most part of the paper is devoted to the spherically symmetric black hole with two horizons -- the charged Reissner-Nordstr\"om (RN) black hole. We  calculate the  entropy and the corresponding temperature of the thermal Hawking radiation using several different approaches. All of them give the same results, which demonstrates the correlations between the inner and outer horizons. Then one of these approaches is applied to the Kerr black hole.

In Sec.\ref{HawkingRadiation} the radiation from the RN black hole is considered using the semiclassical 
tunneling approach.\cite{Wilczek2000,Srinivasan1999,Volovik1999,Akhmedov2006,Vanzo2011,Jannes2011}
When both inner and outer horizons are taken into account, one obtains that the temperature of radiation does not coincide with the conventional Hawking temperature related to the outer horizon. The modified Hawking temperature does not depend on the charge $Q$, and thus is the same as in case of the Schwarzschild black hole with the same mass $M$, i.e. $\tilde T_{\rm H} = \frac{1}{8\pi M}$.

In Sec.\ref{Interpretation} it is shown that this result can be written as the product of the thermal factors, determined by the Hawking temperatures of two horizons: $\exp{\left(-\frac{E}{\tilde T_{\rm H} }\right)} =
\exp{\left(-\frac{E}{T_+}\right)}  \times \exp{\left(-\frac{E}{T_-}\right)}$. This demonstrates that the Hawking radiation in the presence of two horizons is the correlated process, which is similar to the phenomenon of cotunneling in the electronic systems, when an electron experiences the coherent sequence of tunneling events:  from an initial to the virtual intermediate states and then to final state state.\cite{Feigelman2005,Glazman2005} In our case the virtual intermediate state is in the region between the two horizons.
The cotunneling also takes place in the process of Hawking radiation of massive particles with energy smaller than mass, $E<m$.\cite{Jannes2011} In this case the particle first tunnels across the horizon, and then tunnels to the detector.

In Sec.\ref{Singular} the entropy is considered in the approach based on the method of the singular coordinate transformations, which was developed in Ref.\cite{Volovik2021a}.
 In this approach, the probability of the macroscopic quantum tunneling between the black hole and white hole is calculated, which allows to find the entropy of both the black and white holes.
 The result for the entropy of RN black hole,  $\tilde S_{\rm RN}= 4\pi M^2$, also does not depend on charge $Q$ and is the same as for neutral black hole with the same mass. This result, which was considered puzzling in Ref.\cite{Volovik2021a}, now has got the support from the modified Hawking temperature:  it is in agreement with the thermodynamic relation $d\tilde S_{\rm RN} =  d(4\pi M^2)=\frac{dM}{\tilde T_{\rm H}}$.

The independence of the entropy on charge $Q$ suggests that the states with different $Q$ can be obtained by the adiabatic transformations. However, this would contradict to the charge conservation. In Sec.\ref{Adiabatic} we consider the adiabatic transformation, which does not contradict to the conservation of the charge $Q$. It corresponds to adiabatic change of the fine structure constant $\alpha$ to its zero value.
When $\alpha$ slowly decreases to zero the two horizons move slowly  with conservation of the charge number $Q$ and energy $M$. 
Finally the inner horizon disappears and the black hole at $\alpha=0$ becomes neutral, in spite of the finite value of charge $Q$. In such slow process the entropy does not change, and is the same as the entropy of the neutral black hole. 

In Sec.\ref{KerrSection} the adiabatic approach of Sec.\ref{Adiabatic} is extended to the Kerr black hole. Here the adiabatic process is considered in which the angular momentum $J$ of the black hole is conserved, while the angular velocity of rotation approaches zero value. This adiabatic process demonstrates that the entropy of the Kerr black hole depends only on mass $M$ of the black hole and does not depend on its angular momentum $J$.

\section{Reissner-Nordstr\"om black hole}

The metric of Reissner-Nordstr\"om black hole
 with the event horizon  at $r_+$  and the inner horizon at $r_-$:
  \begin{equation}
ds^2= - dt^2 a(r)   + dr^2\frac{1}{a(r)} +r^2 d\Omega^2.
\label{staticRN1}
\end{equation}
where
 \begin{equation}
a(r) =\frac{(r-r_-)(r-r_+)}{r^2} \,.
\label{staticRN}
\end{equation}
The positions of horizons  are expressed in terms of the mass $M$ and charge $Q$ of the black hole  ($G=c=1$):
 \begin{equation}
r_+r_-= Q^2\,\,, \,\, r_+ + r_-= 2M\,.
\label{r+r-}
\end{equation}

The conventional entropy of RN black hole is typically related to the area of the outer horizon:
 \begin{equation}
S_{\rm RN}(r_+)= \pi r_+^2= \pi \left( M + \sqrt{M^2 - Q^2}\right)^2
\label{entropyRN}
\end{equation}
However, in Ref.\cite{Volovik2021a} the different result was obtained, i.e. the entropy does not depend on $Q$: 
 \begin{equation}
\tilde S_{\rm RN}= 4\pi M^2 \,.
\label{entropyRNtilde}
\end{equation}
This entropy can be written in the form:
\begin{equation}
\tilde S_{\rm RN}= \left( \sqrt{S_+} + \sqrt{S_-}\right)^2 \,,
\label{entropyRNtilde2}
\end{equation}
where $S_+$ and $S_-$ are the entropies of the external and inner horizons correspondingly.
The extra term $2 \sqrt{S_+  S_-}$ to the sum of the entropies, $S_+ + S_-$, comes from the correlations between the two horizons.

Eq.(\ref{entropyRNtilde2}) also describes the entropy of  Schwarzschild black hole 
in the de Sitter space-time (SdS), which was obtained in Ref.\cite{Shankaranarayanan2003}  (see Eq.(68) therein). In the SdS case, $S_+$ and $S_-$ are the entropies of the  cosmological horizon and the black hole horizon correspondingly.
The general concept of temperature in multi-horizon spacetimes with application to SdS metric is discussed in Ref. \cite{Padmanabhan2007}.

Similar result takes place for the total entropy of the Kerr black hole (including that of the inner
horizon). It is determined by the mass $M$ and does not depend on the black hole angular momentum $J$.\cite{Okamoto1992}

First we explain such behaviour of the entropy for the spherically symmetric RN black hole, We shall use several different approaches. Then  in Sec.\ref{KerrSection} we apply one of these approaches to the Kerr black hole.

\section{Hawking radiation from two horizons}
\label{HawkingRadiation}

The Eq.(\ref{entropyRNtilde}) can be obtained from the temperature of the thermal Hawking radiation, which can be found 
using the method of the semiclassical quantum tunneling.\cite{Wilczek2000,Srinivasan1999,Volovik1999,Akhmedov2006,Vanzo2011}
 Now we have two horizons, and  thus the Hawking radiation temperature can be modified. For RN black hole the PG metric is not well defined.  In the semiclassical method the process of Hawking radiation is typically studied using the Painleve-Gullstrand (PG) metric.\cite{Painleve,Gullstrand} However, the existence of the inner horizon does not allow for the well defined  PG metric to exist,\cite{Faraoni2020} and its extension is required. The proper extension of the PG  metric is
obtained by the following coordinate transformation:\cite{Volovik2003}
 \begin{equation}
dt \rightarrow dt \pm fdr  \,\,,\, f=\frac{\sqrt{2Mr}}{\sqrt{r^2 + Q^2}}\frac{r^2}{(r-r_-)(r-r_+)}\,.
\label{CoordinateTrnaformationRN}
\end{equation}
This leads to the black hole and white hole states without singularities at horizons:
 \begin{equation}
ds^2= - dt^2\left( 1 +\frac{Q^2}{r^2}\right)+ \frac{1}{1 +\frac{Q^2}{r^2}} (dr\pm v dt)^2+r^2 d\Omega^2\,,
\label{PG_RN}
\end{equation}
where the shift velocity is:
 \begin{equation}
v^2 = \frac{2M}{r} \left( 1 +\frac{Q^2}{r^2}\right)\,.
\label{velocity_RN}
\end{equation}
In this extension of the PG coordinates, the shift velocity is real for all $r$. This is distinct from the conventional  (non-modified) PG coordinates,\cite{Hamilton2008,Zubkov2019} where the shift velocity becomes imaginary for $r<r_0= r_+r_-/(r_+ +r_-)$.

 The tunneling trajectory for massless particle 
can be found from the energy spectrum of a particle. The latter is determined by  the contravariant metric $g^{\mu\nu}$, which is inverse to the PG metric $g_{\mu\nu}$ in Eq.(\ref{PG_RN}):\cite{Volovik2003}
 \begin{equation}
g^{\mu\nu}p_\mu p_\nu=0 \,\, \rightarrow  E = p_r v(r) \pm p_r \left(1 + \frac {Q^2}{r^2}\right)\,.
\label{InverseMetric}
\end{equation}
Here $p_r$ is the radial momentum, and $E=p_0$. 

The probability of the tunneling process is given by the exponent of  imaginary part of the action along the tunneling trajectory, ${\rm Im} \int p_r(r,E) dr$,
where the trajectory $p_r(r,E)$ is 
 \begin{equation}
p_r(r,E) =\frac{E}{ v(r) -\left(1 + \frac {Q^2}{r^2}\right)}\,.
\label{pr}
\end{equation}
In the system with two horizons, the imaginary part of the action is produced by both poles. The contribution of two poles gives the following probability of the Hawking radiation:
 \begin{eqnarray}
P= \exp\left( - 4\pi E \left(  \frac{r_+^2}{r_+ -r_-} - \frac{r_-^2}{r_+ -r_-} \right)  \right) =
\label{P1}
\\
=  \exp\left( - 4\pi E (r_+ +r_-) \right).
\label{P2}
\end{eqnarray}
This corresponds to the thermal radiation characterized by the Hawking temperature, which is  modified by the influence of the inner horizon:
 \begin{equation}
\tilde T_{\rm H} = \frac{1}{4\pi (r_+ +r_-)}= \frac{1}{8\pi M} \,.
\label{HawkingTtilde1}
\end{equation}
The modified Hawking temperature is fully determined by the mass $M$ of the RN black hole, i.e. it does not depend
on the charge $Q$.

The modified temperature of Hawking radiation agrees with the entropy of the black hole in Eq.(\ref{entropyRNtilde}), which also does not depend on $Q$. According to thermodynamic equation one has:
 \begin{equation}
d\tilde S_{\rm RN} =  d(4\pi M^2)=\frac{dM}{\tilde T_{\rm H}} \,.
\label{ThermodynamicsTtilde}
\end{equation}

The Hawking temperature $\tilde T_{\rm H}$ remains constant, when the black hole approaches the extremal limit. That is why
contrary to the proposal in Ref. \cite{Hod2021}, the radiation quanta emitted by the near-extremal  RN black hole cannot transform it into the naked singularity, and thus the Penrose cosmic censorship conjecture is not violated.

Anyway, it would be interesting to consider the limit of extremal black hole.
When $r_+ - r_-  \rightarrow 0$, one should take into account the non-zero imaginary part in the spectrum of radiating particles. This leads to the finite range of the charge $Q$ near the extremal limit, in which the black hole entropy changes from $4\pi M^2$ to zero.

 \section{Interpretation in terms of Hawking temperatures at two horizons}
 \label{Interpretation}
 
It is interesting that Eq.(\ref{P1}) can be interpreted in terms of the conventional Hawking temperatures at two horizons.  Eq.(\ref{P1})  can be written as  the product of two thermal factors:
  \begin{equation}
P=P_+P_-=\exp{\left(-\frac{E}{T_+}\right)}  \times \exp{\left(-\frac{E}{T_-}\right)}  \,.
\label{cotunneling}
\end{equation}
Here $T_+$ and $T_-$ are the conventional Hawking temperatures at two horizons, which are determined by the gravity at the horizons:\cite{ZhaiLiu2010}
\begin{eqnarray}
T_+= \frac{1}{4\pi} a'(r_+)= \frac{1}{4\pi} \frac{r_+ -r_-}{r_+^2} \,,
\label{T+}
\\
T_-= \frac{1}{4\pi} a'(r_-)= -\frac{1}{4\pi}  \frac{r_+ - r_-}{r_-^2} \,.
\label{T-}
\end{eqnarray}
The  temperature $T_+$ determines the rate of the tunneling from the region $r_-<r<r_+$ to $r>r_+$, while the negative  temperature $T_-$ determines the occupation number of these particles near the inner horizon.

Altogether the correlated effect of two horizons leads to the modified Hawking temperature in Eq.(\ref{HawkingTtilde1}):
  \begin{equation}
P=P_+P_- = \exp{\left(-\frac{E}{\tilde T_{\rm H}} \right)} \,.
\label{cotunneling}
\end{equation}
This correlated process is similar to the phenomenon of cotunneling in the electronic systems -- the coherent process of electron tunneling  from an initial to final state via virtual intermediate state.\cite{Feigelman2005,Glazman2005} In our case the virtual intermediate state is in the region between the two horizons.

\section{Entropy in approach based on the singular coordinate transformations}
\label{Singular}

The same expressions for the entropy and the Hawking temperature of the NR black hole can be obtained using the approach in Ref.\cite{Volovik2021a}.   In this approach the macroscopic quantum tunneling is considered from the PG RN black hole in Eq.(\ref{PG_RN}) to the static state in Eq.(\ref{staticRN1}). The latter is viewed as intermediate state between the black hole and white holes states in Eq.(\ref{PG_RN}). Assuming that the intermediate state  has zero entropy, one obtains the entropy of the black and white holes.  

The probability of the quantum tunneling from the RN black hole to its static RN partner in Eq.(\ref{staticRN}) with the same mass $M$ and charge $Q$ is determined by singularities in the coordinate transformation (\ref{CoordinateTrnaformationRN}), which change the action $\int Mdt$ of the black hole. 
The  coordinate transformation in Eq.(\ref{CoordinateTrnaformationRN}) has two singularities (at two horizons). They produce the  change in the action, which imaginary part determines the macroscopic quantum tunneling from the PG RN black hole to its static partner:
\begin{eqnarray}
 P_{\text{BH-RN}\rightarrow \text{static-RN}}  =  \exp{ \left(-2\,{\rm Im} \int M d\tilde t \right)} =
 \label{P_RN1}
 \\
 = \exp{ \left(-2\,{\rm Im} \int M(dt  + f(r)dr ) \right)} = 
 \label{P_RN2}
 \\
 = \exp{ \left(-2M\,{\rm Im}  \int f(r)dr  \right)} = 
 \label{P_RN3}
 \\
 = \exp{ \left(-\pi (r_+ + r_-)^2/G\right)} =   \exp{ \left(-4\pi M^2G\right)} \,.
\label{P_RN4}
 \end{eqnarray}
The probability of the tunneling appears to be the same as for the neutral black hole with the same mass $M$. 

The probability of the quantum tunneling between the PG black hole and its static partner with the same mass can be considered as the quantum fluctuation, which is determined by the difference in the entropy of the two objects.\cite{Landau_Lifshitz} Then  under the natural assumption that the entropy of the intermediate fully static object is zero, the Eq.(\ref{P_RN4}) gives the entropy of the RN black hole in agreement with  Eq.(\ref{entropyRNtilde}):
\begin{equation}
S_\text{BH-RN}(M,Q)=S_\text{BH}(M,0) =4\pi M^2 \,.
\label{entropy_RN}
\end{equation}

The  consideration of the further tunneling -- from the static RN state to the RN white hole state -- provides the negative entropy for the Reissner-Nordstr\"om white hole:\cite{Volovik2021a}
 \begin{equation}
S_\text{WH-RN}(M,Q)=- S_\text{BH-RN}(M,Q)= -4\pi M^2
 \,.
\label{entropy_RN2}
\end{equation}
The quantum tunneling from the RN black hole to the RN white hole with the same mass  via the intermediate static state is determined by the following exponent:
\begin{equation}
\exp{\left(S_\text{WH-RN}(M,Q)- S_\text{BH-RN}(M,Q)\right)}=\exp{\left( -8\pi M^2\right)}
 \,.
\label{BlackToWhite}
\end{equation}
It represents one of the many routes of the transformation from the black hole to the white hole.\cite{Barcelo2014,Barcelo2017,Rovelli2018,Rovelli2019,Rovelli2018b,Uzan2020,Uzan2020b,Uzan2020c,Rovelli2021}
It is interesting that while the formation of the gravitational white hole has extremely small probability, in hydrodynamic analog gravity the white hole horizon is easily formed as the  hydraulic jump in the flowing shallow water.\cite{Volovik2005a,Volovik2006a,JannesRousseaux2012,Berry2018,Berry2019} 

All the considerations above demonstrated that the factor $(1+Q^2/r^2)$ in Eqs.(\ref{PG_RN}) and (\ref{velocity_RN})  is irrelevant for the entropy and Hawking temperature of the RN black hole. This suggests that this factor can be removed by some adiabatic process, at which the entropy does not change. We consider the possibility of such process in the following section.

 \section{Entropy from adiabatic transformation}
\label{Adiabatic}

The independence of $\tilde S_{\rm RN}$ and $\tilde T_{\rm H}$ on the charge $Q$ may suggest that the RN black hole  can be obtained from the neutral BH by adiabatic process, at which the entropy does not change. But this would contradict to the conservation of charge $Q$ in the adiabatic process. However, let us remember that we used the equations, in which the fine structure constant 
$\alpha$ was hidden. With the parameter $\alpha$ involved one has the following dependence on $\alpha$:
 \begin{equation}
r_+r_-= \alpha q^2\,\,, \,\,  r_\pm =  M \pm  \sqrt{M^2 - \alpha q^2} \,,
\label{entropyRNalpha}
\end{equation}
Here $Q^2=\alpha q^2$, where $q$ is conserved number, when measured in terms of the charge of electron, and the parameter
 $\alpha$ -- the fine structure constant -- is the property of the quantum vacuum, which can be varied by changing the parameters of the Standard Model. 
The gravity with varying $\alpha$  is the particular case of gravity with nonlinear electrodynamics (discussion of the latter see e.g. in Refs. \cite{Odintsov2017,Kruglov2020,Amirabi2021} and references therein).

Now we can adiabatically transform the RN black hole  by slow change of this parameter $\alpha$ to zero at fixed mass $M$ and fixed quantum number $q$.
Since in the adiabatic process the entropy does not change, it is the same as for $\alpha=0$, and thus it does not depend on $\alpha$ and $Q$ giving rise to Eq.(\ref{entropyRNtilde}):
\begin{equation}
\tilde S_{\rm RN}(\alpha >0,M)= \tilde S_{\rm RN}(\alpha =0,M)=4\pi M^2 \,.
\label{entropyRNtildePositive}
\end{equation}

At first glance, it looks impossible to fix all three quantities (mass $M$, charge $q$ and entropy $S$).
However, the RN black hole has extra degrees of freedom in the thermodynamics, which come from the inner horizon. 
When $\alpha$ slowly decreases to zero,  the two horizons in Eq.(\ref{entropyRNalpha}) move slowly  with conservation of the charge number $q$ and energy $M$ until the inner horizon disappears.  In such slow process the entropy does not change, and thus it is the same as the entropy of the neutral black hole without inner horizon. 

The independence of the entropy on $\alpha$ may solve the problem of the quantization rule, which is inspired by the string theory.\cite{Horowitz1996,Gibbons2011}
This quantization conjecture requires the square root of a rational number value of the fine structure constant, while the latter is determined by the ultraviolet cut-off.\cite{Visser2012,Visser2013}

It would be interesting to consider the extension to the negative $\alpha$, where only single horizon remains.

\section{Entropy of Kerr black hole from adiabatic transformation}
\label{KerrSection}

In Sec. \ref{Adiabatic} we considered the adiabatic transformation of  the RN black hole to the neutral black hole  by the slow change of the fine structure parameter $\alpha$ to zero at fixed mass $M$ and quantum number $Q$. In case of the Kerr black hole we need the transformation to the black hole to the non-rotating black hole with the same mass $M$ and the angular momentum $J$. The non-rotating limit corresponds to zero value of the angular velocity $\Omega \rightarrow 0$ at fixed $M$ and $J$. Now we shall use two parameters, one is the Planck energy scale $E_P$ which is kept fixed, and another one is the Planck constant $\hbar$, which we change. The variable $\hbar$ emerges for example in effective theory as the parameter of Minkowski metric.\cite{Volovik2009}
The non-rotating limit with fixed $J$ is reached at $\hbar \rightarrow \infty$.

The Kerr metric has the form:
\begin{eqnarray}
ds^2 = - \frac{\Delta}{\rho^2}(dt - a \sin^2\theta d\phi)^2 + \frac{\rho^2}{\Delta} dr^2 +\rho^2 d\theta^2 +
\nonumber
\\
+\frac{\sin^2\theta}{\rho^2} [adt - (r^2+a^2)d\phi]^2
\,,
\label{KerrMetric}
\end{eqnarray}
where
\begin{eqnarray}
 \rho^2(r.\theta)=r^2+a^2 \cos^2\theta\,,\,\,a=\frac{J}{M} \,,
\label{parameter_a}
\\
\Delta(r) = r^2 + a^2 - 2Mr \frac{\hbar}{E_P^2} \,.
\label{parameter_hbar}
\end{eqnarray}
Let us now fix the parameters $J$, $M$ and $E_P$ and adiabatically change the parameter $\hbar$.
This change influences only the function $\Delta$ in Eq.(\ref{parameter_hbar}), while the parameter $a=J/M$ remains fixed, since it does not depend on $\hbar$.

The equation $\Delta(r_\pm)=0$ determines the positions of the inner and outer horizons. At fixed $J$, $M$ and $E_P$
and at $\hbar \rightarrow \infty$ they behave as: 
\begin{equation}
r_+\sim 2\frac{\hbar M} {E_P^2} \,\,, \,\, r_- \sim \frac{J^2 E_P^2}{2M^3 \hbar}\,.
\label{InnerKerr}
\end{equation}
While the outer horizon grows, the inner horizon shrinks, and in the limit $\hbar \rightarrow \infty$, the entropy is produced only by the outer horizon:
\begin{equation}
S\rightarrow \pi r_+^2 \frac{E_P^2}{\hbar^2} = 4\pi \frac{M^2}{E_P^2}\,.
\label{EntropyKerr}
\end{equation}

Since the process of transformation of the rotating Kerr black hole to the "non-rotating" Kerr black hole with $j=J/\hbar \rightarrow 0$ is adiabatic, it demonstrates that that the entropy of the Kerr black hole depends only on mass $M$:
\begin{equation}
S_{\rm Kerr}(J,M)= S_{\rm Kerr}(J=0,M)= 4\pi \frac{M^2}{E_P^2}\,.
\label{EntropyKerr2}
\end{equation}
Eq.(\ref{EntropyKerr2}) is similar to Eq.(\ref{entropyRNtildePositive}) for the RN black hole. The entropy of Kerr black hole does not depend on the classical angular momentum $J$ and on its quantum number $j$.

Eqs. (\ref{entropyRNtildePositive}) and (\ref{EntropyKerr2}) can be extended to the Kerr–Newman black hole. Using the adiabatic process with $\alpha \rightarrow 0$ and $\hbar \rightarrow \infty$, one obtains that the total entropy of the Kerr–Newman black hole is fully determined by its mass:
\begin{equation}
S_{\rm KN}(J,Q,M)= S(J=0,Q=0,M)= 4\pi \frac{M^2}{E_P^2}\,.
\label{EntropyKN}
\end{equation}
Note that independence of the entropy  on $J$ and $Q$ is supported by the consideration of quasinormal modes,\cite{Medved2008,Vagenas2008} which suggests that the quantization of the black hole entropy does not depend on $J$ and $Q$:
\begin{equation}
\Delta S= 8\pi \frac{M \Delta M}{E_P^2}=2\pi n \,.
\label{Quantization}
\end{equation}

Let us also mention that the extremal Kerr black hole, in which the two horizons merge, can be also reached in the adiabatic process with fixed $M$ and $J$. This takes place when  $\hbar \rightarrow J (E_P^2/M^2)$. The same is with the extremal RN black hole, which can be reached by the proper deformation of the fine structure parameter $\alpha$, as we discussed in Sec. \ref{HawkingRadiation}. 

 \section{Conclusion}

We considered the thermodynamics of the black holes with two horizons. We calculated the  entropy of Reissner-Nordstr\"om black hole and the corresponding temperature of the thermal Hawking radiation
using several different approaches. These are: 

(i) The method of semiclassical tunneling for calculation of the Hawking temperature in Sec.\ref{HawkingRadiation}. 

(ii) The cotunneling  mechanism -- the coherent sequence of tunneling at two horizons, each determined by the corresponding Hawking temperature in Sec. \ref{Interpretation}. 

(iii) The macroscopic quantum tunneling for calculation of the transition rate from the RN black hole to the RN white hole in Sec.\ref{Singular}. 

(iv) The adiabatic change of the fine structure constant $\alpha$ to zero and this the adiabatic transformation from the RN black hole to the Schwarzschild black hole in Sec.\ref{Adiabatic}. 

All of the approaches give the same result. The correlations between the inner and outer horizons
lead to the total entropy and to the temperature of Hawking radiation, which depend only on mass $M$ of the black hole and do not depend on the black hole charge $Q$.

The adiabatic approach (iv) has been also applied to the Kerr and Kerr–Newman black holes, which demonstrated that the total entropy of these black holes depends only on mass $M$ and does not depend on the black hole angular momentum $J$ and charge $Q$.

  {\bf Acknowledgements}. This work has been supported by the European Research Council (ERC) under the European Union's Horizon 2020 research and innovation programme (Grant Agreement No. 694248).

\end{document}